\documentclass[11pt,aps,pra,onecolumn,floatfix,amsmath,amssymb,notitlepage,superscriptaddress,tightenlines]{revtex4-1}

\usepackage{graphicx}

\usepackage[letterpaper, total={16.5cm, 25.0cm}]{geometry}

\usepackage[caption=false]{subfig}
\usepackage{amsmath}	
\usepackage{amssymb}    
\usepackage[pdftex]{hyperref}
\begin{document}



\title{Numerical computation of dynamical Schwinger-like pair production in graphene}

\author{Fran\c{c}ois Fillion-Gourdeau}
\email{francois.fillion@emt.inrs.ca}
\affiliation{Universit\'{e} du Qu\'{e}bec, INRS-\'{E}nergie, Mat\'{e}riaux et T\'{e}l\'{e}communications, Varennes, Qu\'{e}bec, Canada J3X 1S2}
\affiliation{Institute for Quantum Computing, University of Waterloo, Waterloo, Ontario, Canada, N2L 3G1}

\author{Philippe Blain}
\affiliation{Universit\'{e} du Qu\'{e}bec, INRS-\'{E}nergie, Mat\'{e}riaux et T\'{e}l\'{e}communications, Varennes, Qu\'{e}bec, Canada J3X 1S2}

\author{Denis Gagnon}
\affiliation{Universit\'{e} du Qu\'{e}bec, INRS-\'{E}nergie, Mat\'{e}riaux et T\'{e}l\'{e}communications, Varennes, Qu\'{e}bec, Canada J3X 1S2}
\affiliation{Institute for Quantum Computing, University of Waterloo, Waterloo, Ontario, Canada, N2L 3G1}

\author{Catherine Lefebvre}
\affiliation{Universit\'{e} du Qu\'{e}bec, INRS-\'{E}nergie, Mat\'{e}riaux et T\'{e}l\'{e}communications, Varennes, Qu\'{e}bec, Canada J3X 1S2}
\affiliation{Institute for Quantum Computing, University of Waterloo, Waterloo, Ontario, Canada, N2L 3G1}

\author{Steve MacLean}
\email{steve.maclean@emt.inrs.ca}
\affiliation{Universit\'{e} du Qu\'{e}bec, INRS-\'{E}nergie, Mat\'{e}riaux et T\'{e}l\'{e}communications, Varennes, Qu\'{e}bec, Canada J3X 1S2}
\affiliation{Institute for Quantum Computing, University of Waterloo, Waterloo, Ontario, Canada, N2L 3G1}

\date{\today}

\begin{abstract}
The density of electron-hole pairs produced in a graphene sample immersed in a homogeneous time-dependent electrical field is evaluated. Because low energy charge carriers in graphene are described by relativistic quantum mechanics, the calculation is performed within the strong field quantum electrodynamics formalism, requiring a solution of the Dirac equation in momentum space. The latter is solved using a split-operator numerical scheme on parallel computers, allowing for the investigation of several field configurations. The strength of the method is illustrated by computing the electron momentum density generated from a realistic laser pulse model. We observe quantum interference patterns reminiscent of Landau-Zener-St\"{u}ckelberg interferometry.  
\end{abstract}

\maketitle


\section{Introduction}

Graphene, a two-dimensional hexagonal array of carbon atoms, can be used as a quantum electrodynamics (QED) simulator because at low energy, quasi-particles in interaction with the lattice are described by a massless Dirac equation \cite{gusynin2007ac}. Close to Dirac points, where the electron momentum $\mathbf{p}$ relative to Dirac point momenta $\mathbf{K}_{\pm}$ obeys $\mathbf{p} \lesssim 100 \ \mbox{eV} \ll |\mathbf{K}_{\pm}| = 3361$ eV, the dispersion relation becomes relativistic-like and is given by $E_{\mathbf{p}} = v_{F}|\mathbf{p}|$, where $v_{F} =  1.093\times 10^{6}$ m/s is the Fermi velocity. In this regime, the quasi-particle dynamics is described by massless reduced quantum electrodynamics RQED$_{3,2}$ where the subscripts denote the dimension of the gauge boson (photon) and fermion (electron), respectively \cite{PhysRevD.86.025005}. In other words, the fermion and boson do not ``live'' in the same number of dimensions. In addition, there are two fermionic species in graphene, associated with the two Dirac points $\mathbf{K}_{\pm}$. The differences and resemblances between graphene RQED$_{3,2}$ and usual QED are summarized in Table \ref{tab:qed_vs_rqed}.

Owing to this analogy, many QED-like phenomena have been studied in graphene such as the Klein paradox \cite{PhysRevLett.102.026807} and the Casimir effect \cite{PhysRevB.84.035446}.  
Recently, Schwinger's pair production mechanism has been considered because monolayer graphene does not have a mass gap and thus, does not suffer from an exponential suppression of the pair production probability \cite{PhysRevD.78.096009}. Electron-positron pair production can be simulated by coupling a graphene sample to an electromagnetic field and by respecting some experimental conditions: thermal and phonon effects should be avoided, a large enough sample should be utilized and the electron-electron coupling $g:= \alpha c/\epsilon v_{F} \ll 1$, where $\alpha \approx 1/137$ is the fine coupling constant and $c$ is the light velocity, should be small enough.

When the electric field interacts with the graphene sample, electron-hole quasi-particle pairs are generated. This is analogous to the generation of electron-positron pairs in QED. As this process can be studied within the strong field RQED$_{3,2}$ formalism, the generation of electron-hole pairs is analogous to Schwinger-like and multiphoton pair production in QED.

\begin{table}
\begin{tabular}{|l|cc|}
\hline
  &QED & Graphene low energy  \\
  && (RQED$_{3,2}$) \\
\hline
Interaction & $e^{+},e^{-}, \gamma$ & Quasi-particles, $\gamma$ \\ 
Fermion dimension & 3D & 2D \\
Photon dimension & 3D & 3D \\
Dispersion relation &$E_{\mathbf{p}} = \pm c \sqrt{\mathbf{p}^{2} + m^{2}}$& $E_{\mathbf{p}} = \pm v_{F}|\mathbf{p}|$\\
Fermion mass & $m_{e^{-}}$ & massless \\
Velocity & Light velocity  & Fermi velocity  \\
&$c \approx 3 \times 10^{8}$ m/s & $v_{F} \approx c/300$\\
Fermionic species & 1  & 2  \\
& (electron) & ($\mathbf{K}^{\pm}$ Dirac points) \\
\hline
\end{tabular}
\caption{Differences between QED and massless reduced QED (RQED$_{3,2}$), the theory that describes graphene quasi-particles in the low energy limit.}
\label{tab:qed_vs_rqed}
\end{table}

In this article, the pair production mechanism in graphene is studied within the strong field  RQED$_{3,2}$ formalism for homogeneous electric fields with general time-dependence. In this formalism, the electron pair density is related to a solution of the Dirac equation coupled to the classical field. Analytic solutions to the Dirac equation exists for some simple cases such as the T-constant field \cite{PhysRevD.86.125022}, the exponential field \cite{1402-4896-90-7-074005} and the Sauter-type field \cite{PhysRevD.82.105026}. In the adiabatic limit, owing to the analogy between the Dirac equation and two-level systems, it is also possible to develop analytical estimates based on the adiabatic perturbation theory or semi-classical techniques \cite{PhysRevLett.104.250402}. In this work, a numerical scheme based on the split-operator method is employed that allows for arbitrary time-dependence and that enables going beyond the above-mentioned analytically solvable cases. As an illustration, we consider the case of a realistic laser pulse.

\section{Pair production in strong field RQED$_{3,2}$}

There exists many equivalent techniques to compute the electron density produced from a strong classical field \cite{Gelis20161,PhysRevD.86.125022,PhysRevD.78.061701,PhysRevD.83.065007,PhysRevA.86.032118}. It is possible to adapt this formalism to RQED$_{3,2}$ allowing for the investigation of similar phenomena in graphene \cite{PhysRevB.92.035401}. The main result of this analysis, when the graphene sample is immersed in a homogeneous electric field, is a relation between the electron-hole surface density $\langle \tilde{n}_{s,a} \rangle$ and a solution of the Dirac equation \cite{PhysRevB.92.035401}:
\begin{eqnarray}
\langle \tilde{n}_{s,a} \rangle = \int  
\frac{d^{2}\mathbf{p}}{2E_{\mathbf{p}}^{\rm out}2E_{\mathbf{p}}^{\rm in}}
 \left|  u^{\mathrm{out} \dagger}_{s,a}(\mathbf{p}) U_{\mathbf{p},a}(t_{f},t_{i}) v^{\mathrm{in}}_{s,a}(-\mathbf{p}) \right|^{2},
\label{eq:pair_prod_homo}
\end{eqnarray}
where $s=\pm 1$ is the electron spin, $a = \mathbf{K}_{\pm}$ characterizes the different Dirac points and
where the evolution operator $U_{\mathbf{p},a}$ evolves the initial wave function $v_{s,a}^{\rm in}$ from the initial time $t_{i}$ to the final time $t_{f}$. Therefore, it gives a solution to the following massless Dirac equation expressed in momentum space  \cite{gusynin2007ac}:
\begin{equation}\label{eq:dirac_eq_mom}
i\partial_{t}\psi_{s,\mathbf{K}_{\pm}}(t,\mathbf{p}) =   \pm v_{F}\boldsymbol{\sigma} \cdot \left[ \mathbf{p}  + e\mathbf{A}(t) \right] \psi_{s,\mathbf{K}_{\pm}}(t,\mathbf{p}),
\end{equation} 
where $e > 0$ is the electric charge, $\mathbf{A}$ is the time-dependent vector potential and $\boldsymbol{\sigma}$ are Pauli matrices. We choose a gauge where $A_{0} = 0$ and thus, any homogeneous electric field can be related to the vector potential by $\mathbf{E}(t) = - \partial_{t}\mathbf{A}(t)$. The electric field vanishes outside the time interval $[t_{i},t_{f}]$. On the other hand, the vector potential can take a constant value (the value depends on the gauge chosen), denoted by $\left. \mathbf{A}(t)\right|_{t \in [-\infty,t_{i}]} = \mathbf{A}^{\mathrm{in}}$ and $\left. \mathbf{A}(t)\right|_{t \in [t_{f},\infty]} = \mathbf{A}^{\mathrm{out}}$.

It is now convenient to introduce the adiabatic free spinors given by
\begin{eqnarray}
\label{eq:free_spin1}
u_{s,\mathbf{K}_{+}}(t,\mathbf{p}) &=& 
\cfrac{1}{\sqrt{E_{\mathbf{p}}(t)}}
\begin{bmatrix}
E_{\mathbf{p}}(t) \\
v_{F}\left[+ P_{x}(t) + iP_{y}(t) \right]
\end{bmatrix},  
v_{s,\mathbf{K}_{+}}(t,-\mathbf{p}) = 
\cfrac{1}{\sqrt{E_{\mathbf{p}}(t)}}
\begin{bmatrix}
v_{F}\left[-P_{x}(t) + iP_{y}(t) \right]\\
E_{\mathbf{p}}(t) 
\end{bmatrix} ,\\
\label{eq:free_spin3}
u_{s,\mathbf{K}_{-}}(t,\mathbf{p}) &=& \cfrac{1}{\sqrt{E_{\mathbf{p}}(t)}}
\begin{bmatrix}
E_{\mathbf{p}}(t) \\
v_{F}\left[-P_{x}(t) - iP_{y}(t)\right]
\end{bmatrix} ,
\label{eq:free_spin4}
v_{s,\mathbf{K}_{-}}(t,-\mathbf{p}) = \cfrac{1}{\sqrt{E_{\mathbf{p}}(t)}}
\begin{bmatrix}
v_{F}\left[+P_{x}(t) - iP_{y}(t) \right]\\
E_{\mathbf{p}}(t) 
\end{bmatrix} ,
\end{eqnarray}
where the kinematic momentum is
\begin{eqnarray}
\mathbf{P}(t) =  \mathbf{p} +e \mathbf{A}(t),
\end{eqnarray}
and where the energy is defined as
\begin{eqnarray}
E_{ \mathbf{p}}(t) &=& v_{F}| \mathbf{P}(t) |.
\end{eqnarray}
The spinors obey the usual property $u^{\dagger}_{s,a}(t,\mathbf{p})v_{s,a}(t,-\mathbf{p}) = 0$. In Eq. \eqref{eq:pair_prod_homo}, free spinors have a subscript $\mathrm{in/out}$, denoting that these spinors are evaluated at times $t_{i}$ and $t_{f}$, respectively ($u_{s,a}^{\mathrm{out}}(\mathbf{p}) := u_{s,a}(t_{f},\mathbf{p})$ and $v_{s,a}^{\mathrm{in}}(\mathbf{p}): = v_{s,a}(t_{i},\mathbf{p})$).

The last undefined quantity in Eq. \eqref{eq:pair_prod_homo} is the evolution operator $U_{\mathbf{p},a}$. This evolution operator should solve Eq. \eqref{eq:dirac_eq_mom} with an initial condition given by the free spinor $v_{s,a}^{\mathrm{in}}(\mathbf{p})$. The formal solution to this initial value problem is given by
\begin{eqnarray}
\psi_{s,a}(t_{f},\mathbf{p}) &=& U_{\mathbf{p},a}(t_{f},t_{i}) v_{s,a}^{\mathrm{in}}(\mathbf{p}),
\end{eqnarray}
where the evolution operator is 
\begin{eqnarray}
\label{eq:time_or}
U_{\mathbf{p},\mathbf{K}_{\pm}}(t_{f},t_{i})
&=& T\exp \left\{\mp i \int_{t_{i}}^{t_{f}}dt \left[  v_{F} \boldsymbol{\sigma} \cdot \mathbf{P}(t)  \right] \right\} ,
\end{eqnarray}
where $T$ stands for the time-ordered operator. The latter is required because the Dirac Hamiltonian does not commute at different times. To compute the electron momentum density, one has to evaluate the effect of the evolution operator on the initial wave function. This is the subject of the next section.

\section{Numerical method for the Dirac equation}

The numerical method employed to evolve the wave function is based on an operator splitting scheme, similar to the ones found in Refs. \cite{PhysRevA.59.604,Mocken2004558,Bauke2011,Mocken2008868}. It is also very similar to the numerical scheme used in \cite{PhysRevA.86.032118} and is a particular version of geometric integrator techniques \cite{mclachlan2002splitting}.
  
The first step is the splitting of the total time interval into $N$ smaller sub-intervals with size $\delta t$. Then, using the semi-group property of the evolution operator defined in Eq. \eqref{eq:time_or}, it is written as
\begin{eqnarray}
\label{eq:ev_op_dt}
U_{\mathbf{p},\mathbf{K}_{\pm}}(t_{f},t_{i}) &=& U_{\mathbf{p},\mathbf{K}_{\pm}}(t_{f},t_{N})U_{\mathbf{p},\mathbf{K}_{\pm}}(t_{N},t_{N-1}) \cdots U_{\mathbf{p},\mathbf{K}_{\pm}}(t_{1},t_{i}).
\end{eqnarray}
The evolution operator can then be expressed in the form \cite{suzuki1993general}
\begin{eqnarray}
\label{eq:sol}
U_{\mathbf{p},\mathbf{K}_{\pm}}(t_{n},t_{n-1})
&=& \exp \bigl\{ i \delta t \left[ \mp v_{F} \boldsymbol{\sigma} \cdot \mathbf{P}(t_{n})  - \mathcal{T} \right] \bigr\} , 
\end{eqnarray}
where $\mathcal{T} = i \overleftarrow{\partial_{t_{n}}}$ is the left time derivative operator. This form of the time-ordered exponential is convenient to derive approximation schemes because it easily lends itself to operator splitting methods. The latter consists in approximating the evolution operator by a sequence of exponentials. A scheme with a third order accuracy is given by the symmetric Strang-like decomposition in the form:
\begin{eqnarray}
\label{eq:approx_time}
U_{\mathbf{p},\mathbf{K}_{\pm}}(t_{n},t_{n-1})
&=& 
e^{- i \frac{\delta t}{2}  \mathcal{T}}
e^{\mp i \delta t  v_{F} \boldsymbol{\sigma} \cdot \mathbf{P}(t_{n}) } 
e^{- i \frac{\delta t}{2}  \mathcal{T}} + O(\delta t^{3}), \\
&=& 
\exp \bigl\{\mp i \delta t  v_{F} \boldsymbol{\sigma} \cdot \mathbf{P}(t_{n+\frac{1}{2}}) \bigr\} 
 + O(\delta t^{3}),
\end{eqnarray}   
where $t_{n+\frac{1}{2}} := t_{n} + \delta t/2$. In principle, the latter can be improved to higher order by using other decompositions \cite{suzuki1993general}.   
Using the properties of Pauli matrices, the exponential can be computed explicitly, yielding
\begin{eqnarray}
U_{\mathbf{p},\mathbf{K}_{\pm}}(t_{n},t_{n-1})
& \approx & \mathbb{I}\cos\left[  \delta t   E_{\mathbf{p}}(t) \right] - i\cfrac{ v_{F}\boldsymbol{\sigma} \cdot \mathbf{P}(t) }{  E_{\mathbf{p}}(t)} \sin\left[  \delta t    E_{\mathbf{p}}(t) \right].
\end{eqnarray}
This expression is a $2 \times 2$ matrix which can be implemented easily on a computer. This completes the description of the numerical scheme. To improve efficiency, the latter is coded and parallelized using a domain decomposition strategy. Because each momentum is independent, the resulting algorithm has a quasi-linear speedup and is scalable to a high number of processors. Moreover, the $L_{2}$-norm of the wave function is conserved because each step of the splitting is a unitary operation. Finally, the time step has to obey $\delta t \lesssim \frac{\pi}{\max E_{\mathbf{p}}}$ to reach convergence. This condition is typical for Dirac equation solvers \cite{Mocken2008868,FillionGourdeau20121403} and guarantees that oscillations in the wave function are resolved.

\section{Numerical results and discussion}

Numerical results obtained from the above-mentioned techniques are now given for a realistic model that simulates counterpropagating linearly polarized laser pulses. This case illustrates the strength of the numerical approach and at the same time, exhibits some interesting physical features related to the phenomenon of quantum interference.

We consider a laser pulse model characterized by an oscillating field superimposed with a carrier envelope. The resulting electric field has the following time dependence:
\begin{eqnarray}
E_{x}(t) = 
\begin{cases}
E_{0}\sin^{2}(\Omega_{\mathrm{rise}}(t-t_{i})) \cos(\omega (t-t_{i}) + \phi) & \mbox{for} \; t \in [t_{i},t_{i} + \pi/2\Omega_{\mathrm{rise}}]\\
E_{0}\cos(\omega (t-t_{i}) + \phi) & \mbox{elsewhere}\\
E_{0}\cos^{2}(\Omega_{\mathrm{fall}}(t-t_{i})) \cos(\omega (t-t_{i}) + \phi) & \mbox{for} \; t \in [t_{f}-\pi/2\Omega_{\mathrm{fall}},t_{f}]
\end{cases},
\label{eq:elec_field}
\end{eqnarray} 
where $\phi$ is the envelope phase, $E_{0}$ is the field strength, $\omega$ is the laser angular frequency and $\Omega_{\mathrm{rise},\mathrm{fall}}$ determine the rise and fall time of the envelope, respectively.

A laser field with a frequency of $\nu = \omega/2\pi = 10$ THz and a field strength of $E_{0} = 1.0 \times 10^{7}$ V/m is considered. These parameters are chosen such that we are in the Schwinger-like (tunnelling) regime where the non-dimensional Keldysh parameter obeys $\gamma = |p_{y}|\omega /eE_{0} \ll 1$. The pulse duration is 5.5 periods and the envelope frequencies are chosen as $\Omega_{\mathrm{rise}} = \Omega_{\mathrm{fall}} = 7.85 \times 10^{12}$ s$^{-1}$. 

\begin{figure}
\includegraphics[width=0.7\textwidth]{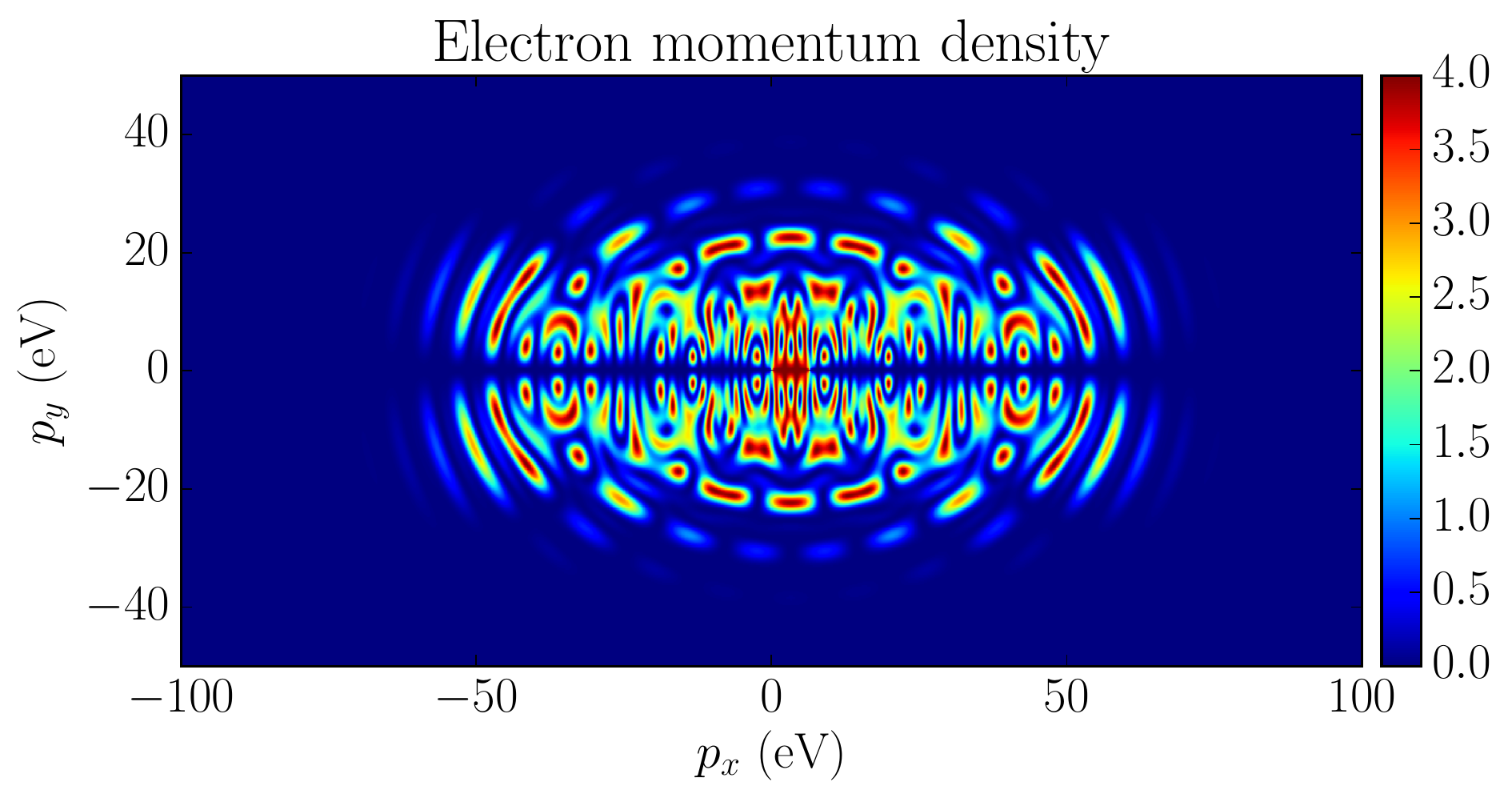}
\caption{Numerical results for the electron momentum density for a laser pulse linearly polarized in the $x$-direction,  having a field strength of $E_{0} = 1.0 \times 10^{7}$ V/m and a frequency of $\nu = 10.0$ THz. The pulse duration is 5.5 periods and the carrier envelope has rise and fall times characterized by the frequencies $\Omega_{\mathrm{rise}} = \Omega_{\mathrm{fall}} = 7.85 \times 10^{12}$ s$^{-1}$ (see Eq. \eqref{eq:elec_field}). }
\label{fig:density_inter}
\end{figure}

The numerical results displayed in Fig. \ref{fig:density_inter} show an intricate ``peak and valley structure'' which is related to time-domain quantum interferences \cite{fillion_interference}. This can be understood by looking at the Dirac equation in Eq. \eqref{eq:dirac_eq_mom}, which is analogous to the equation describing a two-level quantum system. As graphene is driven periodically by the electric field, the adiabatic energies for positive and negative energy states $\pm E_{\mathbf{p}}(t)$ cross in the complex time plane. At these crossings, a nonadabatic transition between positive and negative energy states takes place and causes the generation of electron-hole pairs. Far from the crossings, the time evolution is adiabatic and transitions are forbidden. In this latter case, pair production does not occur but the states accumulate a phase. Because the accumulated phase differs for positive and negative energy states, when they recombine coherently at the next nonadiabatic transition, constructive or destructive interferences occur. This is a realization of Landau-Zener-St\"{u}ckelberg interferometry \cite{Shevchenko20101} in graphene.    

\section{Conclusion}

We showed a parallel numerical scheme that can be employed to evaluate Schwinger-like pair production in graphene from the interaction of a sample with an homogeneous time-dependent electric field. This numerical technique can accommodate any electric field time dependence. As an example and to illustrate the capability of the numerical scheme, it was used to compute the electron momentum density from a realistic short laser pulse. The latter could be generated experimentally by using two counterpropagating laser beams. It should be noted that such realistic field configurations cannot be treated analytically.

The numerical results displayed some interesting features. In particular, the electron momentum density generated from the laser pulse showed interference patterns due to Landau-Zener-St\"{u}ckelberg interferometry.   

Other field configurations could also be studied. For example, it is possible to modify the polarization and study pair production in circularly polarized beams \cite{PhysRevB.92.035401}. Other electric field time-dependence can also be considered, allowing for the investigation of other regimes such as the multiphoton regime \cite{PhysRevB.92.035401}.



\bibliographystyle{apsrev}

\bibliography{bibliography}

\end{document}